\documentstyle[12pt]{article}
\begin{document}

\title{Nuclear matter with a Bose condensate of dibaryons in 
relativistic Hartree approximation}

\author{Amand Faessler$^a$, A. J. Buchmann$^a$, and M. I. 
Krivoruchenko$^{a,b}$  \\
{\small $^a${\it Institut f\"ur Theoretische Physik, Universit\"at
T\"ubingen, Auf der Morgenstelle 14 }}\\
{\small {\it D-72076 T\"ubingen, Germany}}\\
{\small $^b${\it Institute for Theoretical and Experimental Physics,
B.Cheremushkinskaya 25}}\\
{\small {\it 117259 Moscow, Russia}}}

\date{}

\maketitle

\begin{abstract}
The Green's functions are constructed and a one-loop calculation is given
for the scalar and vector densities and for the equation of state of nuclear
matter with a Bose condensate of dibaryons. This is the lowest approximation
in the loop expansion of quantum hadrodynamics, sufficient to account for the
presence of dibaryons not in the condensate in the heterophase
nucleon-dibaryon matter. It leads to a finite effective
nucleon mass and remains consistent with increasing the density.
\vspace{0.5 cm}

\end{abstract}

\newpage 

\section{{\bf Introduction} }

The prospect of observing the
long-lived H-particle predicted in 1977 by Jaffe \cite{Jaf} stimulated
considerable activities in the experimental search for dibaryons \cite{San}.
Although no positive evidence has been observed yet, future experimental
studies may elucidate the nature of this interesting possibility. 
Non-strange dibaryons which have a small width due to zero coupling to the $
NN$-channel are promising candidates for experimental searches \cite{Mul}. A
method for searching exotic dibaryon resonances in the double
proton-proton bremsstrahlung reaction is discussed in Ref. \cite{Ger}. The
narrow peaks observed by Bilger {\it et al.} \cite{Bil} and Clement {\it et
al.} \cite{HCl} in the pion double charge exchange (DCE) reactions on nuclei
have been interpreted by Martemyanov and Schepkin \cite{Mar} as evidence for
the existence of a narrow d' dibaryon with a mass of $2060$ MeV. Recent
experiments at TRIUMF (Vancouver) and CELSIUS (Uppsala) seem to support the
existence of the d' \cite{Mey}. Recently, some indications for a d$_1$(1920)
dibaryon have also been found \cite{Khr}.

The existence of dibaryons when settled reliably by experiments should have
important implications for the properties of nuclear matter. The chemical 
potential of nucleons increases with the density. When it becomes higher 
than half of the chemical potential of dibaryons, the production of dibaryons 
becomes energetically favourable. Dibaryons are Bose particles. In nuclear 
matter they form a Bose condensate. This interesting phenomenon is studied by
Baldin {\it et al}. \cite{Bal} and Chizov {\it et al}. \cite{Chi} by
including dibaryon interactions through a van der Waals volume
correction. A model for nuclear matter with an admixture of dibaryons with
the short-range nuclear forces approximated by a $\delta $-function-like
pseudopotential is discussed in Ref. \cite{Mro}. The relativistic mean-field
theory (MFT) with dibaryons was analysed recently by Faessler {\it et al}. 
\cite{MFT}-\cite{MFT finite T}. An exactly solvable one-dimensional model
for the dibaryon condensation has been found by Buchmann {\it et al.} \cite
{Buc}. Heterophase nucleon-dibaryon matter (HNDM) can be formed in
interiors of neutron stars \cite{Kri}-\cite{Oli} and in heavy-ion collisions 
\cite{MFT finite T}. The physics of heterophase substances is reviewed 
in Ref. \cite{Shu}.

In previous works \cite{Bal}-\cite{MFT finite T}, \cite{Kri}-\cite{Shu},
it was assumed that at zero temperature dibaryons in nuclear matter are all
in the condensate. In Bogoliubov's model of a dilute interacting Bose
gas \cite{Bog}, a fraction of Bose particles is out of the condensate \cite
{Hua,Abr}. This fraction increases with the density and has an important
effect on various thermodynamical quantities.

MFT constitutes the basis of quantum hadrodynamics (QHD) 
\cite{Wal}-\cite{Sero}. In a loop expansion of QHD, it corresponds to the
lowest order term (no loops). In MFT all dibaryons are in the condensate. 
Dibaryons that are not in the condensate appear first 
in a one-loop calculation of the
scalar and vector densities and of the equation of state (EOS). This
approximation is identical to the relativistic Hartree approximation (RHA)
for normal nuclear matter.
It is sufficient to account for the existence of non-condensate dibaryons. 
In this paper, we construct Green's functions of HNDM and give a
one-loop calculation of the scalar and vector densities and the EOS of the
binary nucleon-dibaryon mixture.

\section{\bf Quantum hadrodynamics with dibaryons}

The dibaryonic extension of QHD is determined by the Lagrangian density 
\begin{equation}
\label{LAG}
\begin{array}{c}
{\cal L}=\bar \Psi (i\partial _\mu \gamma _\mu -m_N-g_\sigma \sigma
-g_\omega \omega _\mu \gamma _\mu )\Psi +\frac 12(\partial _\mu \sigma
)^2-\frac 12m_\sigma ^2\sigma ^2-\frac 14F_{\mu \nu }^2 \\ +\frac 12m_\omega
^2\omega _\mu ^2-\frac 12\lambda (\partial _\mu \omega _\mu )^2+(\partial
_\mu -ih_\omega \omega _\mu )\varphi ^{*}(\partial _\mu +ih_\omega \omega
_\mu )\varphi -(m_D+h_\sigma \sigma )^2\varphi ^{*}\varphi .
\end{array}
\end{equation}
Here, $\Psi $ is the nucleon field, $\omega _\mu $ and $\sigma $ are fields
of the $\omega $- and $\sigma $-mesons, $F_{\mu \nu }=\partial _\nu \omega
_\mu -\partial _\mu \omega _\nu $ is the field strength tensor, and 
$\varphi $ is the dibaryon
isoscalar-scalar (or isoscalar-pseudoscalar) field. Furthermore, $m_\omega \ $
and $m_\sigma $ are the $\omega $- and $\sigma $-meson masses and  $
g_\omega $, $g_\sigma $, $h_\omega $, $h_\sigma $ are coupling constants of
the $\omega $- and $\sigma $-mesons with nucleons ($g$) and
dibaryons ($h$). The $\omega $-meson field is described by 
Stueckelberg's equation \cite{IZ}. The limit $\lambda \rightarrow 0$
recovers the Proca equation. The MFT results \cite{MFT}-\cite{MFT finite T}
do not depend on the value of $\lambda $ in Eq.(\ref{LAG}).

The field operators can be expanded in $c$-numbers and operator parts: $
\omega _\mu =g_{\mu 0}\omega _c+\hat \omega _\mu ,$ $\sigma =\sigma _c+\hat
\sigma ,$ $\varphi =\varphi _c+\hat \varphi ,$ and $\varphi ^{*}=\varphi
_c^{*}+\hat \varphi ^{*}$. The $\sigma $-meson mean field determines the
effective nucleon and dibaryon masses in the medium: 
\begin{equation}
\label{ENM}
\begin{array}{c}
m_N^{*}=m_N+g_\sigma \sigma _c, \\ 
m_D^{*}=m_D+h_\sigma \sigma _c. 
\end{array}
\end{equation}

The nucleon scalar and vector densities are defined by the expectation
values 
\begin{equation}
\label{NS}\rho _{NS}=<\bar \Psi \Psi >, 
\end{equation}
\begin{equation}
\label{NV}\rho _{NV}=<\bar \Psi \gamma _0\Psi >. 
\end{equation}
The chemical potential equals $\mu _N=$ $\mu _N^{*}+g_\omega \omega _c$ and $%
\mu _N=$ $-\mu _N^{*}+g_\omega \omega _c$, respectively,\ for nucleons
and for nucleon holes in the Dirac sea. Here, $\mu _N^{*}=+\sqrt{%
m_N^{*2}+k_F^2}$ is the nucleon Fermi energy and $k_F$ is the Fermi
wave number.

The dibaryon scalar density is defined as an additional source of 
the $\sigma $-meson mean field 
\begin{equation}
\label{DS}2m_D^{*}\rho _{DS}=<2(m_D+h_\sigma \sigma )\varphi ^{*}\varphi >. 
\end{equation}
The condensate dibaryons give a contribution $\rho _{DS}^c=|\varphi _c|^2$
to the scalar density$.$ The time evolution of the condensate 
field $\varphi_c $is determined by the chemical potential $\mu _D$: 
\begin{equation}
\label{V}\varphi _c(t)=e^{-i\mu _Dt}\sqrt{\rho _{DS}^c}. 
\end{equation}

To eliminate the time dependence from the condensate parts of the $\varphi $
fields, we pass to the $\mu $-representation for dibaryons. This can be
done by the substitution $\varphi \rightarrow \varphi e^{i\mu _Dt}$ and $
\varphi ^{*}\rightarrow \varphi ^{*}e^{-i\mu _Dt}$. The dibaryon vector
density is equal to the timelike component of the expectation value of the
dibaryon current 
\begin{equation}
\label{VI}<j_\mu ^D>=<\varphi ^{*}(2\mu _D-2h_\omega \omega +i\stackrel{%
\leftrightarrow }{\partial })_\mu \varphi >. 
\end{equation}
Here, $(\mu _D)_\mu =(\mu _D,{\bf 0})$ in the rest frame of the substance.
The condensate dibaryons give a contribution $\rho _{DV}^c=2\mu _D^{*}\rho
_{DS}^c$ to the vector density with $\mu _D=\mu _D^{*}+h_\omega \omega _c$.

The self-consistency equation for the effective nucleon mass is:
\begin{equation}
\label{SC}m_N^{*}=m_N-\frac{g_\sigma }{m_\sigma ^2}(g_\sigma \rho
_{NS}+h_\sigma 2m_D^{*}\rho _{DS}). 
\end{equation}
The $\omega $-meson mean field is determined from the equation 
$$m_\omega
^2\omega _c=g_\omega \rho _{NV}+h_\omega \rho _{DV}.$$

\section{\bf Green's functions of heterophase nucleon-dibaryon matter}

The model of Eq.(\ref{LAG}) is analysed within MFT at zero temperature 
in Refs. \cite{MFT,MFT with GF}. To go beyond MFT, it is necessary to 
construct Green's functions of the bosons 
\begin{equation}
\label{VII}iD^{AB}(x^{\prime }-x)=<T\hat A(x^{\prime })\hat B(x)>
\end{equation}
with $A$, $B=$ $\sigma $, $\omega _\mu $, $\varphi $, $\varphi ^{*}.$ For a
one-loop calculation, it is sufficient to construct the 
Green's functions in the
no-loop approximation. Above the critical density for formation of
dibaryons, the $\sigma $- and $\omega $-mesons can be absorbed by the
dibaryons in the condensate. As a result, the dibaryons leave the
condensate and propagate as normal particles. These processes occur at 
tree level, and the Green`s functions in the no-loop approximation
describe the effect of $\sigma -\omega -\varphi -\varphi ^{*}$ mixing.

The Green`s functions can be determined self-consistently by solving 
a system of Gorkov-Dyson equations. Multiplying the equation of motion 
for the $\sigma $ field corresponding to the Lagrangian 
density (\ref{LAG}) by $\hat \sigma $
and taking the time-ordered product, we find the average value of the
equation over the ground state

$$
\begin{array}{c}
(-\Box -m_\sigma ^2)<T\sigma (1)\hat \sigma (2)>=\delta ^4(1,2)+g_\sigma
<T\bar \Psi (1)\Psi (1)\hat \sigma (2)> \\ 
+2h_\sigma <T(m_D+h_\sigma \sigma (1))\varphi ^{*}(1)\varphi (1)\hat \sigma
(2)>. 
\end{array}
$$
Taking into account the second-order terms with respect to the operator
fields (one-loop approximation), one gets 
$$
\begin{array}{c}
(-\Box -m_\sigma ^2)<T\hat \sigma (1)\hat \sigma (2)>=\delta
^4(1,2)+2h_\sigma ^2\rho _{DS}^c<T\hat \sigma (1)\hat \sigma (2)> \\ 
+2m_D^{*}h_\sigma \sqrt{\rho _{DS}^c}(<T\hat \varphi (1)\hat \sigma
(2)>+<T\hat \varphi ^{*}(1)\hat \sigma (2)>). 
\end{array}
$$
In the momentum representation, the equation takes the form 
\begin{equation}
\label{A1}D^{\sigma \sigma }(k)=\tilde D^{\sigma \sigma }(k)+\tilde
D^{\sigma \sigma }(k)2m_D^{*}h_\sigma \sqrt{\rho _{DS}^c}(D^{\varphi \sigma
}(k)+D^{\varphi ^{*}\sigma }(k)). 
\end{equation}

The equations for the other Green's functions can be obtained in similar way:
\begin{equation}
\label{A2}D_\mu ^{\sigma \omega }(k)=\tilde D^{\sigma \sigma
}(k)2m_D^{*}h_\sigma \sqrt{\rho _{DS}^c}(D_\mu ^{\varphi \omega }(k)+D_\mu
^{\varphi ^{*}\omega }(k)), 
\end{equation}
\begin{equation}
\label{A3}D^{\sigma \varphi }(k)=\tilde D^{\sigma \sigma
}(k)2m_D^{*}h_\sigma \sqrt{\rho _{DS}^c}(D^{\varphi ^{*}\varphi
}(k)+D^{\varphi \varphi }(k)), 
\end{equation}
\begin{equation}
\label{A4}D^{\sigma \varphi ^{*}}(k)=\tilde D^{\sigma \sigma
}(k)2m_D^{*}h_\sigma \sqrt{\rho _{DS}^c}(D^{\varphi ^{*}\varphi
^{*}}(k)+D^{\varphi \varphi ^{*}}(k)), 
\end{equation}
\begin{equation}
\label{A5}D_{\mu \nu }^{\omega \omega }(k)=\tilde D_{\mu \nu }^{\omega
\omega }(k)+\tilde D_{\mu \tau }^{\omega \omega }(k)h_\omega \sqrt{\rho
_{DS}^c}[(2\mu _D^{*}+k)_\tau D_\nu ^{\varphi \omega }(k)+(2\mu
_D^{*}-k)_\tau D_\nu ^{\varphi ^{*}\omega }(k)], 
\end{equation}
\begin{equation}
\label{A6}D_\mu ^{\omega \varphi }(k)=\tilde D_{\mu \tau }^{\omega \omega
}(k)h_\omega \sqrt{\rho _{DS}^c}[(2\mu _D^{*}+k)_\tau D^{\varphi \varphi
}(k)+(2\mu _D^{*}-k)_\tau D^{\varphi ^{*}\varphi }(k)], 
\end{equation}
\begin{equation}
\label{A7}D_\mu ^{\omega \varphi ^{*}}(k)=\tilde D_{\mu \tau }^{\omega
\omega }(k)h_\omega \sqrt{\rho _{DS}^c}[(2\mu _D^{*}+k)_\tau D^{\varphi
\varphi ^{*}}(k)+(2\mu _D^{*}-k)_\tau D^{\varphi ^{*}\varphi ^{*}}(k)], 
\end{equation}
\begin{equation}
\label{A8}D^{\varphi \varphi ^{*}}(k)=\tilde D^{\varphi \varphi
^{*}}(k)+\tilde D^{\varphi \varphi ^{*}}(k)[h_\omega \sqrt{\rho _{DS}^c}%
(2\mu _D^{*}+k)_\tau D_\tau ^{\omega \varphi ^{*}}(k)+2m_D^{*}h_\sigma \sqrt{%
\rho _{DS}^c}D^{\sigma \varphi ^{*}}(k)], 
\end{equation}
\begin{equation}
\label{A9}D^{\varphi \varphi ^{*}}(k)=\tilde D^{\varphi \varphi
^{*}}(k)[h_\omega \sqrt{\rho _{DS}^c}(2\mu _D^{*}+k)_\tau D_\tau ^{\omega
\varphi ^{*}}(k)+2m_D^{*}h_\sigma \sqrt{\rho _{DS}^c}D^{\sigma \varphi
^{}}(k)], 
\end{equation}
\begin{equation}
\label{A10}D^{\varphi ^{*}\varphi ^{*}}(k)=\tilde D^{\varphi ^{*}\varphi
}(k)[h_\omega \sqrt{\rho _{DS}^c}(2\mu _D^{*}-k)_\tau D_\tau ^{\omega
\varphi ^{*}}(k)+2m_D^{*}h_\sigma \sqrt{\rho _{DS}^c}D^{\sigma \varphi
^{*}}(k)]. 
\end{equation}
Here,
$$
\tilde D^{\sigma \sigma }(k)=\frac 1{k^2-\tilde m_\sigma ^2}, 
$$
$$
\tilde D_{\mu \nu }^{\omega \omega }(k)=\frac{-g_{\mu \nu }+k_\mu k_\nu
/\tilde m_\omega ^2}{k^2-\tilde m_\omega ^2}-\frac{k_\mu k_\nu /\tilde
m_\omega ^2}{k^2-\tilde m^2}, 
$$
$$
\tilde D^{\varphi \varphi ^{*}}(k)=\frac 1{(k+\mu _D^{*})^2-m_D^{*2}} 
$$
are the $\sigma $- and $\omega $-meson and the dibaryon MFT propagators. The
effective $\sigma $- and $\omega $-meson masses are
\begin{equation}
\label{MEM}
\begin{array}{c}
\tilde m_\sigma ^2=m_\sigma ^2+2h_\sigma ^2\rho _{DS}^c, \\ 
\tilde m_\omega ^2=m_\omega ^2+2h_\omega ^2\rho _{DS}^c. 
\end{array}
\end{equation}
The mass of the effective scalar in Stueckelberg's equation 
is given by $\tilde m^2=\tilde m_\omega ^2/\lambda .$

Equations (\ref{A1}) - (\ref{A10}) constitute a closed system of
equations, which allows us to determine selfconsistently the boson propagators
including $\omega -\sigma -\varphi ^{*}-\varphi $ mixing. 
For more details see Ref. \cite{MFT with GF}.

The scalar and vector dibaryon densities are expressed in terms of the
dibaryon Green's functions 
\begin{equation}
\label{DGF}
\begin{array}{c}
D^{\varphi \varphi ^{*}}(k)=(\tilde D^{\varphi ^{*}\varphi }(k)^{-1}-\Sigma
^{\varphi ^{*}\varphi }(k))/\Xi (k), \\ 
D^{\varphi \varphi }(k)=\Sigma ^{\varphi \varphi }(k)/\Xi (k), \\ 
D^{\varphi ^{*}\varphi ^{*}}(k)=\Sigma ^{\varphi ^{*}\varphi ^{*}}(k)/\Xi
(k). 
\end{array}
\end{equation}
The denominator $\Xi (k)$ and the self-energy operators are given by the
expressions 
\begin{equation}
\label{DSE}
\begin{array}{c}
\Xi (k)=(\tilde D^{\varphi \varphi ^{*}}(k)^{-1}-\Sigma ^{\varphi \varphi
^{*}}(k))(\tilde D^{\varphi ^{*}\varphi }(k)^{-1}-\Sigma ^{\varphi
^{*}\varphi }(k))-\Sigma ^{\varphi ^{*}\varphi ^{*}}(k)\Sigma ^{\varphi
\varphi }(k), \\ 
\Sigma ^{\varphi \varphi ^{*}}(k)=\Sigma ^{\varphi ^{*}\varphi
}(-k)=(h_\omega 
\sqrt{\rho _{DS}^c})^2(2\mu _D^{*}+k)_\mu \tilde D_{\mu \nu }^{\omega \omega
}(k)(2\mu _D^{*}+k)_\nu \\ +(2m_D^{*}h_\sigma 
\sqrt{\rho _{DS}^c})^2\tilde D^{\sigma \sigma }(k), \\ \Sigma ^{\varphi
\varphi }(k)=\Sigma ^{\varphi ^{*}\varphi ^{*}}(-k)=(h_\omega 
\sqrt{\rho _{DS}^c})^2(2\mu _D^{*}+k)_\mu \tilde D_{\mu \nu }^{\omega \omega
}(k)(2\mu _D^{*}-k)_\nu \/ \\ +(2m_D^{*}h_\sigma \sqrt{\rho _{DS}^c}
)^2\tilde D^{\sigma \sigma }(k). 
\end{array}
\end{equation}
The structure of the dibaryon Green's functions is identical to the
structure of the Green's functions of fermions in superconductors \cite
{Abr,Kon}.

>From Eqs.(\ref{DSE}) we see that the longitudinal component of the $
\omega $-meson propagator contributes to the self-energy operators of the
dibaryon Green`s functions. The model is renormalizable for any 
finite value of $\lambda .$ In the limit $\lambda \rightarrow 0$ the finite
renormalizable expressions become infinite. 
Now, we put $\lambda =1\ $in order to obtain the $\omega $-meson propagator
used in usual nuclear matter QHD calculations\cite{Chin,Sero}. 
In this case, the QHD results for ordinary nuclear
matter are reproduced at low densities where there is no dibaryon
condensate.

The chemical potential of dibaryons is determined from the relativistic
extension of the Hugenholtz-Pines relation \cite{Hug} 
\begin{equation}
\label{HP}\mu _D^{*2}-m_D^{*2}=\Sigma ^{\varphi \varphi ^{*}}(0)-\Sigma
^{\varphi \varphi }(0). 
\end{equation}
A relation of such a kind is necessary in order to get a pole in the 
dibaryon Green's
functions at $\omega ={\bf k}=0$ and to guarantee the 
existence of sound in the medium. Eq.(\ref{HP}) gives 
\begin{equation}
\label{HPR}\mu _D^{*}=m_D^{*} 
\end{equation}
in agreement with MFT \cite{MFT,MFT with GF}.

\section{\bf One-loop scalar and vector densities and the EOS}

The one-loop expression for the nucleon scalar density is calculated from
Eq.(\ref{NS}). After renormalization, one gets \cite{Chin,Sero} 
\begin{equation}
\label{DENS}\rho _{NS}=\rho _{NS}^c+\rho _{NS}^{\prime }
\end{equation}
with 
\begin{equation}
\label{DENS0}\rho _{NS}^c=\gamma _N\int \frac{d^{}{\bf k}}{(2\pi )^3}\frac{%
m_N^{*}}{\sqrt{m_N^{*2}+{\bf k}^2}}\theta (k_F-|{\bf k}|),{\rm \;}\;\;\;\;
\end{equation}
\begin{equation}
\label{DENS'}\;\rho _{NS}^{\prime }=-4m_N^3\zeta (m_N^{*}/m_N).
\end{equation}
Here, $\gamma _N$ is the statistical factor ($\gamma _N=2$ in neutron matter
and $\gamma _N=4$ in nuclear matter) and
$$
4\pi ^2\zeta (x)=x^3\ln x+1-x-\frac 52(1-x)^2+\frac{11}6(1-x)^3. 
$$

The one-loop expression for the dibaryon scalar density defined by Eq.(\ref
{DS}) can be written in the form 
\begin{equation}
\label{XIII}2m_D^{*}\rho _{DS}=2m_D^{*}\rho _{DS}^c+2m_D^{*}<\hat \varphi
^{*}\hat \varphi >+2h_\sigma \sqrt{\rho _{DS}^c}(<\hat \sigma \hat \varphi
>+<\hat \sigma \hat \varphi ^{*}>). 
\end{equation}
We extract the zero order contribution with respect to the 
dibaryon condensate: 
\begin{equation}
\label{XIV}\rho _{DS}^{\prime }=i\int \frac{d^4k}{(2\pi )^4}\tilde
D^{\varphi \varphi ^{*}}(k). 
\end{equation}
After renormalization, we obtain 
\begin{equation}
\label{DEDS'}2m_D^{*}\rho _{DS}^{\prime }=m_D^3\zeta (m_D^{*}/m_D). 
\end{equation}

The total dibaryon scalar density has the form

\begin{equation}
\label{DEDS}\rho _{DS}=\rho _{DS}^c+\rho _{DS}^{\prime }+\rho _{DS}^{\prime
\prime } 
\end{equation}
with 
\begin{equation}
\label{DEDS''}\rho _{DS}^{\prime \prime }=i\int \frac{d^4k}{(2\pi )^4}
\{D^{\varphi \varphi ^{*}}(k)-\tilde D^{\varphi \varphi ^{*}}(k)+4(h_\sigma 
\sqrt{\rho _{DS}^c})^2\tilde D^{\sigma \sigma }(k)(D^{\varphi \varphi
^{*}}(k)+D^{\varphi \varphi }(k))\}. 
\end{equation}
This integral diverges. The renormalized expression is given by 
\begin{equation}
\label{XVIII}2m_D^{*}\rho _{DS}^{\prime \prime }=\frac 1{8\pi
^2}\int_0^{+\infty }\! dk_E \, k_E^3
\lambda _{DS}^{ren}(k_E,m_D^{*},\rho _{DS}^c) 
\end{equation}
where 
\begin{equation}
\label{XIX}
\begin{array}{c}
\lambda _{DS}^{ren}(k_E,m_D^{*},\rho _{DS}^c)=\lambda _{DS}(k_E,m_D^{*},\rho
_{DS}^c)-\lambda _{DS}(k_E,m_D^{*},0) \\
-\rho _{DS}^c\frac{\partial \lambda _{DS}(k_E,m_D,0)}{\partial \rho _{DS}^c}
-(m_D^{*}-m_D)\rho _{DS}^c\frac{\partial ^2\lambda _{DS}(k_E,m_D,0)}{
\partial m_D^{*}\partial \rho _{DS}^c}. 
\end{array}
\end{equation}
We have passed here into the Euclidean space. The density $\lambda
_{DS}(k_E,m_D^{*},\rho _{DS}^c)$ has the form 
\begin{equation}
\label{XX}\lambda _{DS}(k_E,m_D^{*},\rho _{DS}^c)=\frac{4(k_E^2E_\omega
+m_D^{*2})z^2+k_E^2(4E_\omega -3-k_E^2/m_D^{*2})}{2m_D^{*}E_\omega k_E^2(%
\sqrt{z^2(z^2+1)}+z^2)} 
\end{equation}
where 
$$
E_\omega =\frac{k_E^2+m_\omega ^2}{k_E^2+\tilde m_\omega ^2}, 
$$
$$
\;z^2=\frac{k_E^2+8m_D^{*2}R}{4m_D^{*2}E_\omega }, 
$$
$$
R=\rho _{DS}^c(\frac{h_\omega ^2}{k_E^2+\tilde m_\omega ^2}-\frac{h_\sigma ^2%
}{k_E^2+\tilde m_\sigma ^2}). 
$$
The counter terms in the Lagrangian density, responsible for the
renormalization 
of the scalar density, are of the form $\delta {\cal L}_s=
(C_1\sigma +C_2\sigma ^2)\varphi ^{*}\varphi .$

The one-loop expression for the dibaryon vector current has the form 
\begin{equation}
\label{CUR}<j_\mu ^D>=2(\mu _D^{*})_\mu \rho _{DS}^c+<\hat \varphi ^{*}(2\mu
_D^{*}+i\stackrel{\leftrightarrow }{\partial })_\mu \hat \varphi >-2h_\omega 
\sqrt{\rho _{DS}^c}(<\hat \omega _\mu \hat \varphi ^{*}>+<\hat \omega _\mu
\hat \varphi >). 
\end{equation}
In the rest frame of the substance, the vector $(\mu _D^{*})_\mu $ is
defined by $(\mu _D^{*})_\mu =(\mu _D^{*},{\bf 0}).$

The nucleon vector density in RHA is the same as in MFT.
The dibaryon vector density can be represented in the form 
\begin{equation}
\label{DEDV}\rho _{DV}=\rho _{DV}^c+\rho _{DV}^{\prime }+\rho _{DV}^{\prime
\prime } 
\end{equation}
with 
\begin{equation}
\label{DEDV'}\rho _{DV}^{\prime }=i\int \frac{d^4k}{(2\pi )^4} 2(\mu
_D^{*}+\omega )\tilde D^{\varphi \varphi ^{*}}(k)\equiv 0, 
\end{equation}
and 
$$
\rho _{DV}^{\prime \prime }=i\int \frac{d^4k}{(2\pi )^4}\{2(\mu
_D^{*}+\omega )(D^{\varphi \varphi ^{*}}(k)-\tilde D^{\varphi \varphi
^{*}}(k)) 
$$
\begin{equation}
\label{DEDV''}+4(h_\omega \sqrt{\rho _{DS}^c})^2\tilde D^{\omega \omega
}(k)(2\mu _D^{*}+\omega )(D^{\varphi \varphi ^{*}}(k)+D^{\varphi \varphi
}(k))\}. 
\end{equation}
In this expression, $\omega $ is the timelike component of the vector $k_\mu 
$.

The renormalized expression for the value $\rho _{DV}^{\prime \prime }$ is
given by 
\begin{equation}
\label{XXV}\rho _{DV}^{\prime \prime }=\frac 1{8\pi ^2}\int_0^{+\infty
} \! dk_E \, k_E^3
\lambda _{DV}^{ren}(k_E,m_D^{*},\rho _{DV}^c) 
\end{equation}
with 
\begin{equation}
\label{XXVI}\lambda _{DV}^{ren}(k_E,m_D^{*},\rho _{DV}^c)=\lambda
_{DV}(k_E,m_D^{*},\rho _{DV}^c)-\lambda _{DV}(k_E,m_D^{*},0)-\rho _{DV}^c%
\frac{\partial \lambda _{DV}(k_E,m_D,0)}{\partial \rho _{DV}^c}. 
\end{equation}
Here, 
\begin{equation}
\label{XXVII}\lambda _{DV}(k_E,m_D^{*},\rho _{DV}^c)=\frac{4(k_E^2E_\omega
+m_D^{*2})z^2+k_E^2(4E_\omega -3)}{2m_D^{*}E_\omega k_E^2(\sqrt{z^2(z^2+1)}%
+z^2)}-\frac 1{m_D^{*}}. 
\end{equation}
The counter term in the Lagrangian density, responsible for the
renormalization of the vector density, has the form $\delta {\cal L}%
_v=C_3(\partial _\mu -ih_\omega \omega _\mu )\varphi ^{*}(\partial _\mu
+ih_\omega \omega _\mu )\varphi .$ The structure of the counter terms shows
that the model is renormalizable to one loop.

Despite the fact that a dibaryon Bose condensate does not exist in 
ordinary nuclei,
dibaryons affect properties of nuclear matter and nuclei
through a Casimir effect described by Eq.(\ref{DEDS'}). This effect is
analysed in Ref. \cite{MFT-RHA}. When $x\rightarrow 1$ 
$\zeta (x)=O((1-x)^4)$, so the vacuum contributions to the scalar density 
of nucleons and
dibaryons, which have opposite signs, are comparable for $4g_\sigma
^4/m_N\approx h_\sigma ^4/m_D$. Dibaryon effects become large for $
h_\sigma /(2g_\sigma )\approx 0.5(4m_D/m_N)^{1/4}\approx 0.84$. 
Explicit calculation \cite{MFT-RHA} shows that the basic properties of nuclear
matter at saturation density can be reproduced for $h_\sigma /(2g_\sigma
)<0.8$, and that the results are not very sensitive to the dibaryon 
mass. The parameter sets of the RHA model for the H-particle, and the
d'-, and d$_1$-dibaryons are shown in Table 1. The 
meson masses $m_\sigma =520$ MeV and $m_\omega
=783 $ MeV and the equilibrium Fermi wave number $k_F=1.3$ fm$^{-1}$ are the
same as in RHA without dibaryons \cite{Chin,Sero}.

The effective nucleon and dibaryon masses, which are  obtained as solutions 
of the self-consistency condition (\ref{SC}) are shown in Fig.1a.
In Fig.1 (b-d) the scalar and vector densities for
nucleons and dibaryons, and the EOS of HNDM are plotted versus the total
baryon number density $\rho _{TV}$ for $h_\sigma /(2g_\sigma )=0.6$ and $
h_\omega /(2g_\omega )=0.8.$ The numerical results are given for $\gamma
_N=4 $ (at low density we start from nuclear matter) and for the d' dibaryon 
mass $m_D=2060$ MeV.
At densities $\rho _{TV}<\rho _1$, the chemical potential of
dibaryons is larger than twice the chemical potentials of nucleons, 
$2\mu _N<\mu_D $, so we have normal nuclear\ matter. In the 
interval $\rho _1<\rho
_{TV}<\rho _2$, nucleons and dibaryons are in chemical equilibrium, i.e. $
2\mu _N=\mu _D$, and the density of nucleons decreases with increasing total
baryon number density. In the interval $\rho _2<\rho _{TV}<\rho _3$, {\rm 
the dibaryon chemical potential is inside of the energy gap for nucleons, $
2(-m_N^{*}+g_\omega \omega _c)<\mu _D<2(m_N^{*}+g_\omega \omega _c)$, and
nuclear matter consists of dibaryons only. } At densities $\rho _3<\rho _{TV}$,
antinucleons come into chemical equilibrium with dibaryons, the relation 
$2(-\mu _N^{*}+g_\omega \omega _c)=\mu _D$ holds true. The density of
antinucleons increases with the total baryon number density. The numerical
values for the densities $\rho _1,$ $\rho _2,$ and $\rho _3$ of the nuclear
and neutron matter are given in Table 1.

The energy density and pressure are calculated from 
\begin{equation}
\label{EOS}\varepsilon =\int \mu _Nd\rho _{TV}\;\;\;\;\;\;{\rm and}%
\;\;\;\;\;\;p=\int \rho _{TV}d\mu _N. 
\end{equation}
In the interval $\rho _2<\rho _{TV}<\rho _3$, one should put $2\mu _N=\mu _D$
by definition.

\section{\bf Conclusion}

In an interacting Bose gas, a fraction of bosons is not in 
the condensate 
\cite{Bog}-\cite{Abr}. Non-condensate dibaryons appear in a
one-loop calculation of quantum hadrodynamics. This approximation
corresponds to the relativistic Hartree approximation for normal 
nuclear matter. We have 
constructed the Green`s functions of heterophase nucleon-dibaryon matter 
and have calculated the 
contribution of non-condensate dibaryons to the scalar and vector densities 
and the equation of state. 
This contribution increases rapidly with the total baryon number density and
becomes dominant as compared to the contribution of condensate dibaryons. 
In mean field 
theory, the effective nucleon mass vanishes with increasing 
density \cite{MFT}-\cite
{MFT finite T}. In relativistic Hartree approximation, the effective nucleon 
mass is finite and positive. In
a loop expansion of quantum hadrodynamics, the relativistic Hartree 
approximation is the lowest approximation sufficient 
to account for the presence of non-condensate dibaryons in heterophase 
nucleon-dibaryon matter, while providing consistent solutions at high 
densities.

The present one-loop calculation 
allows (i) to consider the structure of neutron stars with dibaryons in
the interiors without restrictions to the total density and (ii) to construct
the phase diagram for the phase transition of normal nuclear matter to 
dibaryon matter and quark matter without restrictions to the densities
and temperatures.

\vspace{5mm}

{\bf Acknowledgments}

The authors are grateful to M. Kirchbach (University of Mainz) and B.
V. Martemyanov (ITEP, Moscow) for useful discussions. M. I. K. acknowledges
the hospitality of the Institute for Theoretical Physics of University of 
T\"ubingen, the Alexander von Humboldt Stiftung for support with a 
Forschungsstipendium and DFG and RFBR for Grant No. Fa-67/20-1.

\newpage\ 

\begin{center}
Table 1\vspace{5mm}

$
\begin{array}{ccccccccc}
{\rm Dibaryon} & {\rm Mass\;(GeV)} & h_\sigma /(2g_\sigma ) & g_\sigma & 
g_\omega & \rho _1/\rho _0 & \rho _2/\rho _0 & \rho _3/\rho _0 & \gamma _N
\\ 
{\rm H} & 2.22 & 0.6 & 8.5277 & 10.4393 & 
\begin{array}{c}
3.62 \\ 
4.86 
\end{array}
& 11 & 33 & 
\begin{array}{c}
2 \\ 
4 
\end{array}
\\ 
{\rm d}^{\prime } & 2.06 & 0.6 & 8.5438 & 10.4687 & 
\begin{array}{c}
2.48 \\ 
3.52 
\end{array}
& 7.9 & 30 & 
\begin{array}{c}
2 \\ 
4 
\end{array}
\\ 
{\rm d}_1 & 1.92 & 0.6 & 8.5606 & 10.4993 & 
\begin{array}{c}
1.31 \\ 
2.23 
\end{array}
& 4.6 & 29 & 
\begin{array}{c}
2 \\ 
4 
\end{array}
\end{array}
$
\end{center}

\newpage\ 

\begin{center}
{\bf Figure captions}\vspace{5mm}
\end{center}

{\bf Fig.1. }(a) The effective nucleon and dibaryon masses versus the total
baryon number density. (b) The scalar density of nucleons of 
Eq.(\ref{DENS}): the contribution of (i) nucleons in the Fermi sphere 
($\rho _{NS}^c$, solid line), (ii) the negative Dirac sea  ($
\rho _{NS}^{\prime }$, dotted line). The individual contributions to
the scalar density of dibaryons according to Eq.(\ref{DEDS}): 
(i) dibaryons in the condensate ($2m_D^{*}\rho _{DS}^c$, solid line), 
(ii) the positive contribution of the zero-point dibaryon 
fluctuations ($2m_D^{*}\rho _{DS}^{\prime }$, dotted line), and (iii)
the term $2m_D^{*}\rho _{DS}^{\prime \prime }$ (dashed line) describing
the contribution of dibaryons not in the condensate. 
The densities are measured in units of the saturation density $\rho _0$ of
nuclear matter. Note the different dimensions of the scalar nucleon
($[fm^{-3}]$) and dibaryon densities ($[fm^{-2}]$). 
 (c) The fraction of nucleons ($\rho _{NV}/\rho _{TV}$),
and dibaryons ($\rho _{DV}^c/\rho _{TV}$) in the condensate (solid lines). 
The fraction of dibaryons not in the condensate 
($\rho _{DV}^{^{\prime \prime }}/\rho _{TV}$) (dashed line). (d) The
energy density and pressure of heterophase nucleon-dibaryon matter
(dashed lines) compared to the RHA without dibaryons (solid lines) versus
the total baryon number density. At densities $\rho _{TV}<\rho _1$ 
normal nuclear\ matter ($\gamma _N=4$) is stable. In the interval $\rho
_1<\rho _{TV}<\rho _2$, nucleons and dibaryons are in chemical
equilibrium. In the interval $\rho _2<\rho _{TV}<\rho _3$, only{\rm \
dibaryons are present in nuclear matter. } At densities $\rho _3<\rho _{TV}$
dibaryons are in  chemical equilibrium with antinucleons.


\newpage 

\begin{thebibliography}{99}
\bibitem{Jaf}  R.L.Jaffe, Phys. Rev. Lett. {\bf 38} (1977) 195.

\bibitem{San}  M.Sano, M.Wakai and H.Bando, Phys. Lett. {\bf B 224} (1989)
359; F.S.Rotondo, Phys. Rev. {\bf D 47} (1993) 3871; A.T.M.Aerts and
C.B.Dover, Phys. Rev. Lett. {\bf 24} (1982) 1752, Phys. Rev. {\bf D 28}
(1982) 1752; {\bf D 29 }(1984) 433; S.Aoki et al., Phys. Rev. Lett. {\bf 65}
(1990) 1729; A.Rusek et al., Phys. Rev. Lett. {\bf 52} (1995) 1580;
K.Shimizu, Rep. Prog. Phys. {\bf 52} (1989) 57.

\bibitem{Mul}  P.J. Mulders, A.T.Aerts and J.J. de Swart, Phys. Rev. {\bf D
21} (1980) 2653; L.A.Kondratyuk, B.V.Martemyanov and M.G.Schepkin, Yad. Fiz. 
{\bf 45} (1987) 1252; L.A.Glozman, A.Buchmann and A.Faessler, J. Phys. {\bf %
G 20} (1994) L49; G.Wagner, L.A.Glozman, A.J.Buchmann and A.Faessler, Nucl.
Phys. {\bf A 594} (1995) 263; A.J.Buchmann, G.Wagner, L.Ya.Glozman and
A.Faessler, Progr. Part. Nucl. Phys. {\bf 36} (1995) 383.

\bibitem{Ger}  S.B.Gerasimov and A.S.Khrykin, Mod. Phys. Lett. {\bf A 8}
(1993) 2457.

\bibitem{Bil}  R.Bilger et al., Phys. Lett. {\bf B 269} (1991) 247;
R.Bilger, H.A.Clement and M.G.Schepkin, Phys. Rev. Lett. {\bf 71} (1993) 42.

\bibitem{HCl}  H.Clement, M.Schepkin, G.J.Wagner and O.Zaboronsky, Phys.
Lett. {\bf B 337} (1994) 43.

\bibitem{Mar}  B.V.Martemyanov and M.G.Schepkin, Pisma v ZhETF, {\bf 53}
(1991) 132.

\bibitem{Mey}  R.Meyer et al., Contribution to PANIC 1996; W.Brodowski et
al., Z. Phys. {\bf C 355} (1996) 5.

\bibitem{Khr}  A.S.Khrykin, $\pi N$ Newsletter, No. {\bf 10} (1995) 67;
Contribution to PANIC-96, USA, May 22-27; V.M.Abazov {\it et al.} Preprint
JINR-E1-96-104, Dubna, 1996.

\bibitem{Bal}  A.M.Baldin {\it et al.}, Dokl. Acad. Sc. USSR {\bf 279}
(1984) 602.

\bibitem{Chi}  A.V.Chizov {\it et al.}, Nucl. Phys. {\bf A 449} (1986) 660.

\bibitem{Mro}  St.Mrowczynski, Phys. Lett. {\bf B 152} (1985) 299.

\bibitem{MFT}  A.Faessler, A.J.Buchmann, M.I.Krivoruchenko and
B.V.Martemyanov, Phys. Lett. {\bf B 391 }(1997){\bf \ } 255.

\bibitem{MFT with GF}  A.Faessler, A.J.Buchmann, M.I.Krivoruchenko and
B.V.Martemyanov (to be published).

\bibitem{MFT finite T}  A.Faessler, M.I.Krivoruchenko and B.V.Martemyanov,
Preprint nucl-th/9706079 (to be published).

\bibitem{Buc}  A.J.Buchmann, A.Faessler and M.I.Krivoruchenko, Ann. Phys.
(N.Y.), {\bf 254} (1997) 109.

\bibitem{Kri}  M.I.Krivoruchenko, JETP Letters {\bf 46} (1987) 3.

\bibitem{Tam}  R.Tamagaki, Progr. Theor. Phys. {\bf 85} (1991) 321.

\bibitem{Oli}  A.Olinto, P.Haensel and J.Frieman, Preprint
FERMILAB-PUB-91-176-A, 1991.

\bibitem{Shu}  A.S.Shumovskij and V.I.Yukalov, Phys. Elem. Part. and At.
Nucl., {\bf 16} (1985) 1274.

\bibitem{Bog}  N.N.Bogoliubov, Izv. Akad. Nauk SSSR, Ser. Fiz. {\bf 11}
(1947) 77.

\bibitem{Hua}  K.Huang, C.N.Yang and J.M.Luttinger, Phys. Rev. {\bf 105}
(1957) 767; K.A.Brueckner and K.Sawada, Phys. Rev. {\bf 106} (1957) 1117.

\bibitem{Abr}  A.A.Abrikosov, L.P.Gorkov and I.E.Dzyaloshinski, {\it Methods
of Quantum Field Theory in Statistical Physics}, 1963 (Prentice-Hall, Inc.,
Englewood Cliffs, New Jersey).

\bibitem{Wal}  J.D.Walecka, Ann. Phys. (N.Y.) {\bf 83} (1974) 491; S.A.Chin
and J.D.Walecka, Phys. Lett. {\bf 52 B} (1974) 24.

\bibitem{Chin}  S.A.Chin, Ann. Phys. (N.Y.) {\bf 108} (1977) 301.

\bibitem{Sero}  B.D.Serot, Rep. Prog. Phys. {\bf 55} (1992) 1855.

\bibitem{IZ}  C.Itzykson and J.-B. Zuber, {\it Quantum Field Theory}, 1980
(McGraw-Hill Book Company, New York e.a.).

\bibitem{Kon}  L.A.Kondratyuk, M.M.Giannini and M.I.Krivoruchenko, Phys.
Lett. {\bf B 269} (1991) 139.

\bibitem{Hug}  N.M.Hugenholtz and D.Pines, Phys. Rev. {\bf 116} (1959) 489.

\bibitem{MFT-RHA}  A.Faessler, A.J.Buchmann, M.I.Krivoruchenko, 
nucl-th/9706080, Phys. Rev. {\bf C56} Vol. 3, Sept. 1997, 
\end{thebibliography}
\end{document}